\documentclass[preprint]{aastex}
\usepackage{graphicx}

\usepackage{natbib}
\usepackage{wrapfig}

\usepackage{graphics}

\usepackage{natbib}

\newcommand{\be}{\begin{equation}}
\newcommand{\ee}{\end{equation}}
\newcommand{\nn}{\mbox{} \nonumber \\ \mbox{} }
\newcommand{\ba}{\begin{eqnarray}}
\newcommand{\ea}{\end{eqnarray}}

\newcommand{\E}{{\vec{E}}}
\newcommand{\B}{{\vec{B}}}

\renewcommand{\div}{{\rm \,div\,}}

\newcommand\eg{\textit{e.g.,\ }}

\newcommand\bfedit\bf

\newcommand{\Bf}{{magnetic field}}
\newcommand{\Bfs}{{magnetic fields}}
\newcommand{\Ef}{{electric  field}}
\newcommand{\Efs}{{electric fields}}

\newcommand{\NS}{neutron star}
\newcommand{\NSs}{{neutron stars}}
\newcommand{\EM}{electromagnetic}
\newcommand{\BH}{{black hole}}
\newcommand{\BHs}{{black holes}}
\newcommand{\Sc}{Schwarzschild}
\newcommand{\ms}{magnetosphere}
\newcommand{\mss}{magnetospheres}

\def\apj{ApJ}                 
\def\apjl{ApJ}                
\def\mnras{MNRAS}             
\def\nat{Nature}              

\begin{document}
\title{Slowly balding  black holes}
\author{Maxim Lyutikov\\
Department of Physics, Purdue University, \\
 525 Northwestern Avenue,
West Lafayette, IN
47907-2036 \\ 
Jonathan C. McKinney\\
Department of Physics, Stanford University, \\
2575 Sand Hill Rd.,
Menlo Park, CA
94025}

\begin{abstract}

  The ``no hair'' theorem, a key result in General Relativity, states
  that an isolated black hole is defined by only three parameters: mass,
  angular momentum, and electric charge;
  this asymptotic state is reached on a  light-crossing time scale. We find that the ``no hair''
  theorem is not formally  applicable for black holes formed from collapse of a
  rotating neutron star.  Rotating neutron stars can self-produce particles via vacuum
  breakdown forming a highly conducting plasma magnetosphere
  such that magnetic field lines are effectively ``frozen-in'' the star both before and during collapse.
  In the limit of no resistivity, this introduces a topological constraint which prohibits the magnetic field
  from sliding off the newly-formed event horizon. As a result, during
  collapse of a neutron star into a black hole, the latter conserves the number of
  magnetic flux tubes $N_B = e \Phi_\infty /( \pi c \hbar)$, where
  $\Phi_\infty \approx 2 \pi^2 B_{NS} R_{NS}^3 /(P_{\rm NS} c)$ is the
  initial magnetic flux through the hemispheres of the progenitor and
  out to infinity.  We test this theoretical result via
  three-dimensional general relativistic plasma simulations of
  rotating black holes that start with a neutron star dipole magnetic field
  with no currents initially present outside the event horizon.  The
  black hole's magnetosphere  subsequently relaxes to the split monopole
  magnetic field geometry with self-generated currents outside the
  event horizon. The dissipation of the resulting equatorial current
  sheet leads to a slow loss of the anchored flux tubes, a process
  that balds the black hole on long resistive time scales rather than
  the short light-crossing time scales expected from the vacuum
  ``no-hair'' theorem.

\end{abstract}

\maketitle

\section{Introduction}

The ``no hair'' theorem \citep[][]{MTW} postulates that all black hole
solutions of the Einstein-Maxwell equations of gravitation and
electromagnetism in general relativity can be completely characterized
by only three externally observable classical parameters: mass,
electric charge, and angular momentum.  The key point in the classical
proof \citep[\eg][]{1972PhRvD...5.2439P} is that the outside medium is a vacuum. In contrast, the
surroundings of astrophysical high energy sources like pulsars and
\BHs\ can rarely be treated as vacuum
\citep{GoldreichJulian,Blandford:1977,1992MNRAS.255...61M}.  The
ubiquitous presence of \Bfs\ combined with high (often relativistic)
velocities produce inductive \Efs\ with electric potential drops high
enough to break the vacuum via various radiative effects (curvature
emission followed by a single photon pair production in \Bf, or
inverse Compton scattering followed by a two photon pair production).
For example, in case of \NSs\ the rotation of the \Bf\ lines frozen
into the crust generates an inductive electric field, which, due to
the high conductivity of the \NS\ interior, induces surface
charges. The \Ef\ of these induced surface charges has a component
parallel to the dipolar \Bf.  These parallel \Efs\ accelerate charges
to the energy $ {\cal E} \sim e B_s R_s ( \Omega R_0/c)^2$, where
$B_s$ and $R_s$ are the surface \Bf, radius of a \NS\ and $\Omega$ is
the angular rotation frequency. The resulting primary beam of leptons
produces a dense secondary plasma via vacuum breakdown. Thus, in case
of \NSs\ the electric charges and currents are self-generated: no
external source is needed.  Rotating black holes can also lead to a
similar vacuum break-down \citep{Blandford:1977}.

In this paper we argue that the ``no hair'' theorem is not applicable
to black holes formed from the collapse of magnetized neutron stars.
In particular, we demonstrate that contrary to the prediction of the
``no hair'' theorem, the collapse of a rotating \NS\ into the \BH\
results in a formation of a long lived self-generated conducting BH
\ms.  This results from the violation of the key assumption of the
``no hair'' theorem, that the outside is vacuum, and allows a \BH\ to
preserve open magnetic flux tubes that initially connect to the \NS\
surface.

\section{Electrodynamics of Neutron Star Collapse}

\subsection{Plasma Electrodynamics: the Constraint of Frozen-in Magnetic Fields}

The electrodynamics of a highly conducting medium is qualitatively
different from vacuum electrodynamics. The key difference is that the
highly conducting plasma quickly shorts out any \Ef\ ($\E$) parallel
to \Bf\ ($\B$) through the induction of electric currents
\citep[][]{Kulsrud}. The condition $\E\cdot \B=0$ introduces a
constraint that the \Bf\ lines are effectively frozen into plasma:
each plasma element is always ``attached'' to a given \Bf\ line.

Neutron stars possess dipolar \Bf\
\citep{1968Natur.219..145P,GoldreichJulian} \citep[this assumption has
recently been verified by direct measurements of the structure of the
pulsar \ms\ in the double pulsar system;][]{lt05}. In addition,
rotation-induced poloidal (i.e. $r$-$\theta$ plane) currents flowing
in the \ms\ lead to opening-up to infinity of some fields lines
originating close to the magnetic polar line. Thus, a \NS\ is
surrounded by self-generated plasma and has field lines that connect
its surface to infinity.

\subsection{Electromagnetically Dominated Media: Force-Free Electrodynamics}

Since the mass density of the self-generated plasma is many orders of
magnitude smaller than the energy density of \Bf, the plasma dynamics
can be treated in the so-called force-free approximation
\citep{Gruzinov99}, when the \EM\ fields evolve due to forces
generated by fields themselves.  But the system is qualitatively
different from vacuum: effectively, massless charges and currents
provide the force balance $\rho_e \E + {\bf j} \times \B =0$ and
ensure the ideal condition $\E\cdot\B=0$. More generally, the MHD
formulation assumes (explicitly) that the second Poincare
electro-magnetic invariant $\E \cdot \B=0$ and (implicitly) that the
first electro-magnetic invariant $B^2 -E^2 > 0$. This means that the
electro-magnetic stress energy tensor can be diagonalized and,
equivalently, that the electric field vanishes in the plasma rest
frame.  This assumption is important since we are interested in the
limit when the matter contribution to the stress energy tensor goes to
zero; the possibility of diagonalization of the electro-magnetic
stress energy tensor distinguishes the force-free plasma and vacuum
electro-magnetic fields (for which such diagonalization is generally
not possible).

The equations of force-free electrodynamics can be derived from
Maxwell equations and the constraint $\E\cdot \B=0$.  This can be done
using general tensorial notation from the general relativistic MHD
formulation in the limit of negligible inertia
\citep{1997PhRvE..56.2181U}.  This offers an advantage that the system
of equations may be set in the form of conservation laws
\citep{2002MNRAS.336..759K}.  A more practically appealing formulation
involves the 3+1 splitting of the equations of general relativity
\citep{ThornMembrane,1989PhRvD..39.2933Z}.  The Maxwell equations in
the stationary Kerr metric then take the form
\ba && 
 \nabla \cdot \E =  4 \pi \rho
 \nn &&
 \nabla \cdot \B=0
 \nn &&
 \nabla  \times ( \alpha \B) = 4 \pi \alpha {\vec{j}} + D_t \E
 \nn &&
 \nabla   \times ( \alpha \E) = - D_t \B ,
 \label{maxw1}
\ea
where $D_t = \partial _t - {\cal L} _{\vec{\beta}} $ is the total time
derivative, including Lie derivative along the velocity of the zero
angular momentum observers (ZAMOs), $ \nabla$ is a covariant
derivative 
and $\alpha$ is the delay function   (in \Sc\ geometry
$\alpha = \sqrt{1-2 M/r}$).  (Relations (\ref{maxw1}) are valid if the  shift  function is divergence-less $\div  \vec{\beta}=0$ and   the metric is time-independent. For a more general formulation see \cite{2004MNRAS.350..427K,2011arXiv1108.3511K}.) Taking the total time derivative of the
constraint $\E \cdot \B =0$ and eliminating $D_t \E$ and $D_t \B$
using Maxwell equations, one arrives at the corresponding Ohm's law in
Kerr metric \citep{2011PhRvD..83f4001L}, generalizing the result of
\cite{Gruzinov99}:
\be
 {\vec{j}} = {\left( \B \cdot \nabla \times (\alpha \B) -\E \cdot \nabla \times (\alpha \E)  \right)  \B  + \alpha (\nabla \cdot \E) \E \times \B \over 4 \pi \alpha B^2} .
\label{GS000}
\ee
Note that this expression does not contain the shift  function $\vec{\beta}$.  
See also section~2.2 in \citet{mckff06b} for a fully covariant derivation
of the 4-current that also does not require time derivatives
and is independent of the shift function.

\subsection{The Black Hole Hair:  the Conserved Poloidal Magnetic Flux}

During \NS\ collapse into a \BH, time dilation near the horizon
and the frame-dragging of the horizon lead to the ``horizon locking''
condition: objects are dragged into corotation with the hole's event
horizon, which has a frequency associated with it of
$
\Omega_H \approx  a c/(2 r_{\rm Sc}) =   9  \times  10^3 {\rm rad s^{-1}} P_{\rm NS,-3}^{-1}
$,
where $a$ is the dimensionless Kerr parameter. The Kerr parameter of
the resulting \BH\ may become fairly large,
$$
a = (4\pi/5) (c/G) \chi R_{NS}^2/[P_{NS} M_{NS}]
= 0.2 (\chi/0.5) (R_{NS}/10 {\rm km})^2 (1 {\rm msec}/P_{NS}) ,
$$
but only for critically rotating \NSs\ and stiff equations of
state. In the above equation $P_{\rm NS}$ is the initial spin of a
\NS, $\chi$ is the central concentration parameter
\citep[][]{2005MNRAS.358..923B} $\chi \approx 0.5$, and we assumed a
standard \NS\ with mass $1.4 $ of the mass of the Sun. A relative
smallness of the Kerr parameter $a$ justifies that the space-time
is approximately \Sc.

Before the onset of the collapse, the electric currents within the
\NS\ create poloidal \Bf. Rotation of the poloidal \Bf\ lines and the
resulting inductive electric field lead to the creation, through
vacuum breakdown, of the conducting plasma and poloidal electric
currents.  The presence of a conducting plasma then imposes a
topological constraint, that the \Bf\ lines which initially were
connecting the \NS\ surface to the infinity must connect the \BH\
horizon to the infinity.

During the collapse, as the surface of a \NS\ approaches the horizon,
the closed \Bf\ lines will be quickly absorbed by the \BH, while the
open field lines (those connecting to infinity) have to remain open by
the frozen-in condition.  Thus, a \BH\ can have only open fields
lines, connecting its horizon to the infinity. There is a well known
solution that satisfies this condition: an exact split monopolar
solution for rotating \ms\ due to \cite{Michel73}; it was generalized
to \Sc\ metrics by \cite{Blandford:1977}. We recently found an exact
non-linear {\it time-dependent } split monopole-type structure of
\mss\ driven by spinning and collapsing \NS\ in \Sc\ geometry
\citep{2011arXiv1104.1091L}. We demonstrated that the collapsing \NS\
enshrouded in a self-generated conducting \ms\ does not allow a quick
release of the \Bfs\ to infinity.

Thus, if a collapsing \BH\ can self-sustain the plasma production in
its \ms, the magnetic field lines that were initially connecting the
\NS\ surface to infinity will connect the \BH\ horizon to the
infinity.  Each hemisphere then keeps the magnetic flux that was
initially connected to the infinity.  For a \NS\ with the surface \Bf\
$B_{NS}$ and the initial pre-collapse radius $R_{NS}$ and period
$P_{\rm NS}$, the magnetic flux through each hemisphere connecting to
infinity is $\Phi_\infty \approx 2 \pi^2 B_{NS} R_{NS}^3 /(P_{\rm NS}
c)$ \citep{GoldreichJulian}. Using quantization of the magnetic flux
\citep[][]{LLV}, this corresponds to a conserved quantum number of
magnetic flux tubes
\be
N_B =  e \Phi_\infty /( \pi c \hbar) = {2 \pi  B_{NS}  e R_{NS} ^3}/({c^2 {\hbar} P_{NS} }) = 10^{41} {B_{NS} \over 10^{12} {\rm G}} \, {P_{NS}  \over 1 {\rm msec}} .
\label{NB}
\ee
This quantum number is the \BH\ ``hair'': an observer at infinity can
measure the corresponding Poynting flux and infer the number $N_B$.

The conserved poloidal magnetic flux (\ref{NB}) implies a \Bf\ on the
horizon of the \BH\
\be
B_{\rm BH} = {\pi \over 4} {c^3 B_{NS} R_{NS}^3 \over (G M)^2 P_{NS}} =
6 \times 10^{11} {\rm G} {B_{NS} \over 10^{12} {\rm G}} \, \left( {P_{NS}  \over 1 {\rm msec}}\right)^{-1} .
\label{BBH}
\ee
We can then verify that the resulting \BH\ will have no problem in
breaking the vacuum: the rotation of the \BH\ leads to the appearance
of the inductive \Ef\ with a total potential drop within the \ms\ of
the order
\be
\Delta \Phi \approx a e B_{\rm BH} R_{\rm BH} = {2 \pi^2 \over 5} \chi { e c^2 \over G^2}
{B_{NS} R_{NS}^5 \over M_{NS}^2 P_{NS}^2} = 10^{19}\, {\rm V} \, {\chi \over 0.5} \, {B_{NS} \over 10^{12} {\rm G}} \, \left( {P_{NS}  \over 1 {\rm msec}}\right)^{-2} .
\label{PP}
\ee
This is sufficiently high to break the vacuum via radiative effects
and produce a highly conducting plasma.  In addition, even in the
relatively weak gravitational field of a \NS , the general
relativistic effects of the rotation of space-time (the Lense-Thirring
precession) dominate the accelerating \Ef\
\citep{1992MNRAS.255...61M}. The Lense-Thirring precession near the
\BH\ also facilitates plasma production \citep{Blandford:1977}.

\section{Numerical Simulations}

We have performed numerical simulations that confirm the basic
principle that the ``no-hair'' theorem and related time-dependent
vacuum simulations are not applicable to a plasma-filled black hole
magnetosphere.  We do not model the process of vacuum breakdown and
the subsequent formation of a plasma-filled magnetosphere.  Instead,
we assume the \NS\ already created a plasma-filled magnetosphere (or
that the black hole self-generates a plasma-filled magnetosphere), and
we assume that the \NS\ has already collapsed to a black hole.
Only once an event horizon has formed would the magnetic
field begin to slip-off the black hole in vacuum, so starting with
an event horizon should be a strong enough test -- one should not
have to follow the collapse of the \NS\ to a black hole as long as a
plasma is present.  The goal of the simulations is to measure the
decay timescale of the magnetic flux threading the event
horizon of the black hole: $\Phi_{\rm EM} = (1/2)\int_S dS |B^r|$ as
integrated over the surface ($S$) of the black hole horizon.  We show
that the magnetic dipole decay seen in vacuum solutions
is avoided or delayed by three effects:
1) presence of plasma and self-generation of toroidal currents ;
2) black hole spin induced poloidal currents ;
and 3) plasma pressure support of current layers generated internally by dissipating currents.
These effects cause the field to avoid vacuum-like decay of the dipole
magnetic field and help support the newly-formed split-monopole \Bf\
against magnetic reconnection.

These GRMHD simulations use the fully conservative, shock-capturing
GRMHD scheme called HARM \citep{gam03,mg04,mck06jfa,mckff06b,nob06}
using Kerr-Schild coordinates in the Kerr metric for a sequence of
spins given by $\{a=0,0.1,0.5,0.9,0.99\}$ (in such models without a
disk, negative spin is not physically distinct from positive
spin). The code includes a number of improvements
\citep{mm07,tch07,tch08,tch09} compared to the original code. This
code is capable of choosing the equations of motion as MHD, force-free
electrodynamics, and vacuum electrodynamics. The code permits full 3D
(no assumed symmetries) simulations as reported in \citet{mb09}.

We perform simulations that either use the force-free or use the fully
energy-conserving MHD equations of motions.  These approximate,
respectively, the limits of radiatively efficient emission and
radiatively inefficient emission once the plasma has been generated.
That is, if the electromagnetic field dominates the rest-mass and
internal energy density over most of the volume outside current
sheets, then the force-free limit corresponds to an instantaneous loss
(such as radiation) of magnetic energy dissipated in current sheets,
while the fully energy-conserving MHD limit without cooling
corresponds to all dissipated energy going into internal+kinetic
energy that remains in the system and sustains the current sheet
against dissipation.  A non-energy-conserving system of equations or
simulation code would be unable to properly follow the energy
conservation process of electromagnetic dissipation within the current
sheet that leads to plasma formation there. The force-free
electrodynamics equations of motion are not solely relied upon because
they are undefined within current sheets and any particular resistive
force-free electrodynamics equations
\citep{LyutikovTear,gruz08,2011arXiv1107.0979L} still leave some
degree of ambiguity in how the resistivity would map onto the full
magnetohydrodynamical (MHD) equations.  For the MHD equations, an
ideal $\gamma=4/3$ gas equation of state is chosen, which can be
considered as mimicking a radiatively inefficient high-energy particle
distribution component generated by the dissipation of the currents
within the reconnecting layer.

The initial condition corresponds to a NS dipole field given by an
orthonormal $\phi$-component of the vector potential of $A_\phi\propto
\sin\theta/r^2$ with no rotation and no $\phi$ component of the
magnetic field present at the initial time.  This initial
magnetic field corresponds to the exterior dipole solution for a
neutron star, but the metric has simply been chosen as a Kerr metric
in Kerr-Schild form.  This forces the black hole alone to produce
any and all currents exterior to the horizon, and so the simulations
represent an even more general test of the horizon locking condition
described in the prior sections.

For MHD models, the magnetosphere is filled with only a low-density
atmosphere of rest-mass density and internal energy density, such that
the electromagnetic energy density exceeds both by factors of ten or
more through-out the simulation.  During the MHD simulations, near the
black hole at the stagnation surface (zero radial velocity, where
plasma must be created) mass-energy is injected in a zero angular
momentum observer (ZAMO) frame to maintain a high magnetization of
roughly $\mu\sim 100$ times more electromagnetic energy than rest-mass
energy. As long as $\mu\gg 1$, the details of how this injection is
done do not affect the results \citep{tch10}.  For force-free
simulations, the condition $B^2-E^2>0$ is enforced as an immediate
loss (e.g. by radiation) of energy-momentum as described in
\citet{mckff06b}.  The initial velocity field is set to be that of a
ZAMO \citep{mg04}.  We have now completely specified the metric,
initial value problem, and equations of motion used.

These models are simulated with increasing resolution to seek
convergence or the trend on results with resolution. The models have
radial resolutions of $N_r=\{64,128,256\}$ cells, $\theta$ resolutions
of $N_\theta=\{32, 64, 128\}$ cells, and $\phi$ resolutions of
$N_\phi=\{1,16,32\}$ cells in order to seek the lowest-order $\phi$
dependence.  The simulation grid resolution is focused on the
magnetosphere near the black hole, while also extending to large radii
to avoid artificial interactions with the outer radial boundary. The
radial grid from $0.9$ times the event horizon radius ($r_{\rm H}$) to
$10^4GM/c^2$ is defined by the exponential $r=0.5+\exp(x_1)$ with
$x_1\propto i/N_r$ for grid element $i$ up to a break radius of
$r=100GM/c^2$ for an easily computable $x_{1,\rm br}$, after which the
grid becomes hyper-exponential such that value of $x_1$ adds to itself
$(x_1-x_{1,\rm br})^2$ for each $i>i_{\rm br}$.  The grid in $\theta$
goes from $0$ to $\pi$ and is uniform.  The grid in $\phi$ goes from
$0$ to $2\pi$ and is also uniform.  The simulations are run for a time
$1000GM/c^3$ in order to reach a quasi-steady state.

At the low resolutions considered for these simulations, dissipation
is enhanced compared to expected at high resolutions. So the
simulations only place lower limits on a measurement of the decay
timescale of the magnetic flux threading the black hole event horizon.
A convergent solution implies the result is independent of any
microscopic resistivity model and only depends upon the
self-consistently generated turbulent resistivity.  A non-convergent
result would imply the microscopic resistivity controls the
dissipation and must be specified.  For such a non-convergent case, we
can estimate the expected resistivity \citep{um11}, such as done in
the next section.

Figure~1 shows the initial and quasi-steady contour plots for
$\Psi=R A_\phi$ for the axisymmetric $256\times 128\times 1$ resolution and
black hole with spin $a=0.99$.  The initial dipole field has collapsed
to essentially a split monopole field with most of the magnetic flux
passing through the horizon instead of reconnecting near the equator.
This was predicted by \cite{2011arXiv1104.1091L}.  All other models at
different spins show qualitatively similar final states, except the 3D
simulations with resolution $256\times 128\times 16$ or $256\times
128\times 32$ show slightly more closed field structures at the
equator beyond the event horizon.

Figure~2 shows the value of magnetic flux $\Phi_{\rm EM}$ vs. time
($t$) for all black hole spins and both the force-free electrodynamics
equations of motion (radiatively efficient regime) and the MHD
equations of motion (radiatively inefficient regime).  At $t=0$, the
black hole rotation begins to drive radial currents into the
magnetosphere.  Then, during the first $t\sim 20GM/c^3$, some portion
of the dipole field is absorbed by the black hole.  For high black
hole spin rates, the amount of lost magnetic flux is found to
correspond to losing those field lines that do not cross the Alfv\'en
surface ($R\sim \Omega_F/c\approx \Omega_H/(2c)$, for field rotation
frequency $\Omega_F$) at late times.  This means that those field
lines that eventually open-up to infinity are prevented from being
absorbed by the black hole by the poloidal currents driven into the
magnetosphere by the rotation of the black hole.  For low black
hole spin rates, such as for $a=0$, the early loss of magnetic flux
is limited by the equatorial toroidal current sheet
that forms over a timescale of $t\sim 25GM/c^3$,
which corresponds to the light-crossing timescale
over a wavelength of $\lambda\approx 24.7GM/c^2$ for
electromagnetic radiation from the decay of a dipole field
for a vacuum \Sc\ black hole \citep{bs03}.  This means that the
decay by vacuum radiation is avoided by toroidal current sheet formation
due to collapse of the dipole field in the presence of a plasma.
Overall, decay of the dipole field is avoided by poloidal and toroidal
currents that spontaneously form due to black hole rotation
and collapse of the dipole field in the presence of a plasma.

For the MHD solutions, over the first $t\sim 50GM/c^3$, the solution
then settles into an quasi-steady state of a roughly constant magnetic
dissipation near the black hole.  The dissipation creates plasma
pressure within the current layer that supports the layer against
magnetic reconnection. The decays are fit well by an exponential decay, where
the decay timescale for the $256\times 128$ resolution for the MHD
models with $a=\{0,0.1,0.5,0.9,0.99\}$ is $\tau\sim
\{100,110,120,400,500\}GM/c^3$, respectively.  As expected for the
discontinuity resolved at the numerical grid scale, the decay
timescale is found to be directly proportional to the resolution
(i.e. $\tau\propto \{N_r,N_\theta,N_\phi\}$).  Otherwise identical 3D
models show larger decay timescales, so even higher resolution 3D
models (i.e. $N_\phi>32$) should not be required to find a lower limit
on the decay timescale.

This figure also shows that the radiative efficiency of the current
layer is crucial to whether the magnetosphere survives for long
periods of time. The force-free solutions continue rapid dissipation
due to a lack of plasma pressure within the current layer. The decay
timescale at this resolution for the force-free models is $\tau\sim
10GM/c^3$ for $a=0$ and $\tau\sim 20GM/c^3$ for $a=0.99$. The
force-free simulations show a decay timescale comparable to vacuum
dipole decay on a black hole, which decays as $\sim (t-19)^{-4}$
starting after only $t\sim 20GM/c^3$ \citep{bs03}.
A force-free simulation code with less dissipation may help avoid
such fast dissipation in the force-free limit (e.g. \citealt{spit06}),
unless the condition $B^2<E^2$ is still manifested within the current layer
because then a causal force-free solution requires dissipation in order
to recover the causal condition $B^2>E^2$.

This figure also shows that non-zero black hole spin induction
of poloidal currents cause an extension of the timescale for dissipation
of the magnetosphere in either the force-free or MHD limits.

\begin{figure}[htb]
\begin{center}$
\begin{array}{cc}
\includegraphics[width=0.5\linewidth]{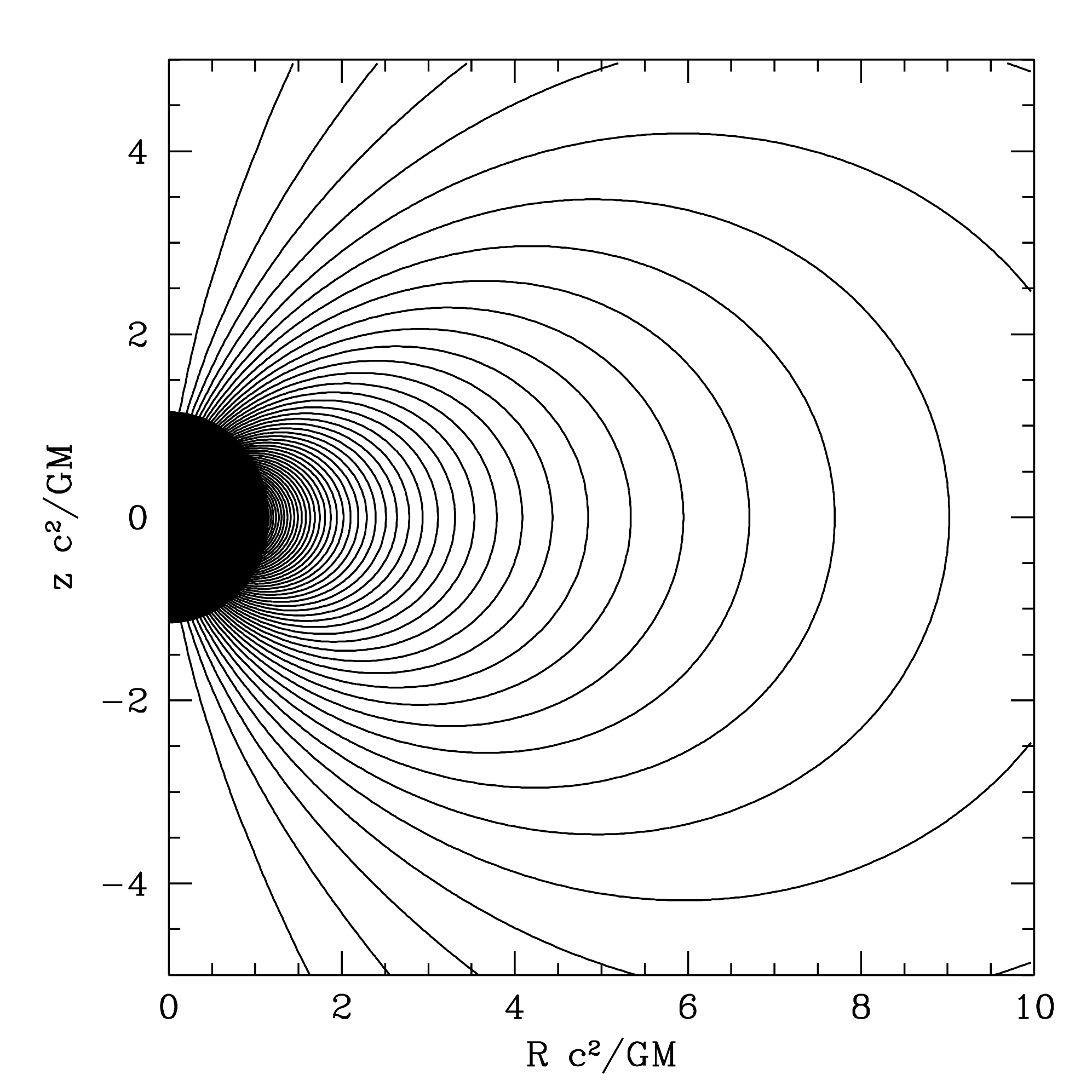}&
\includegraphics[width=0.5\linewidth]{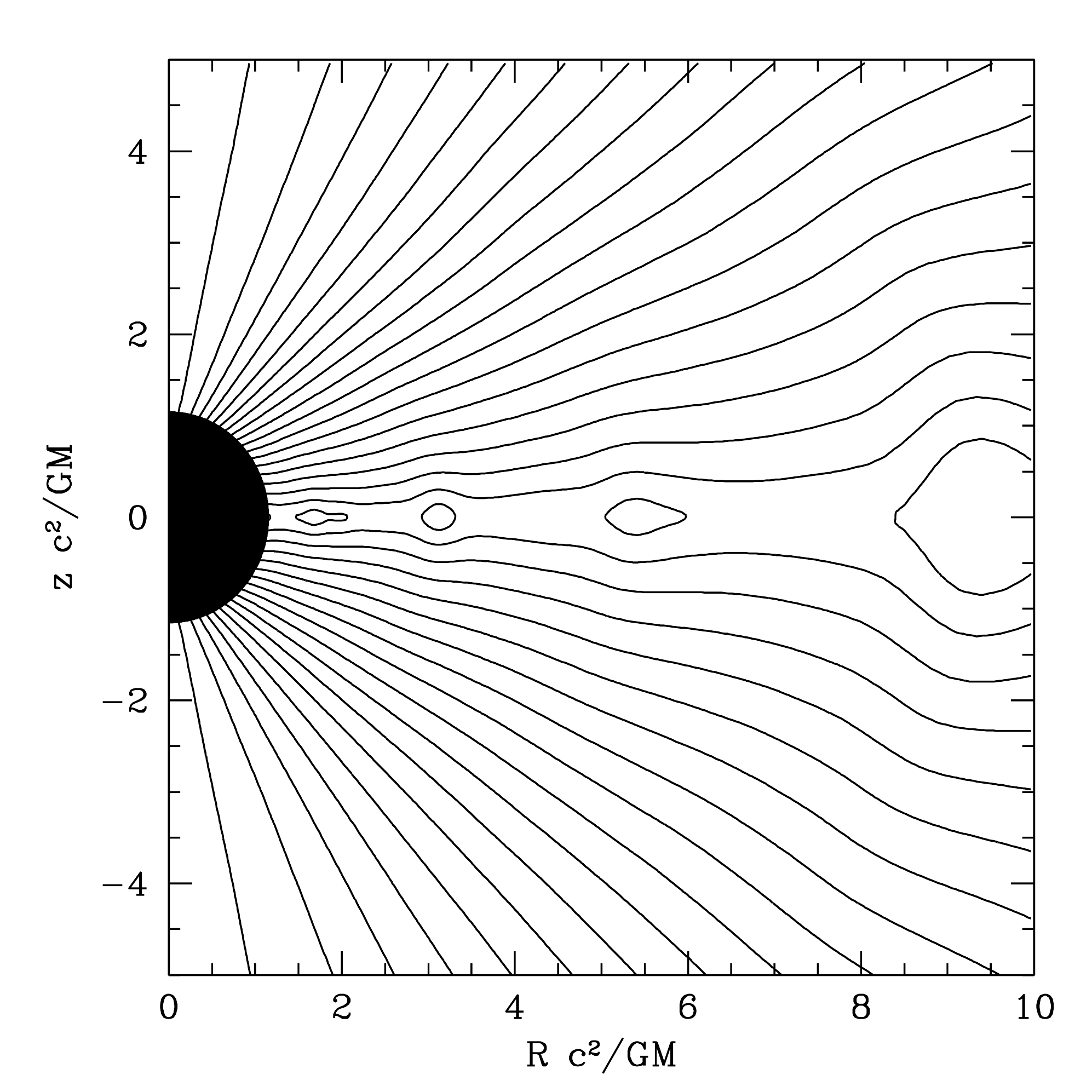}

\end{array}$
\caption{ A contour plot of the magnetic flux ($\Psi=R A_\phi$)
  showing the inner (cylindrical radius) $R<10GM/c^2$ for the MHD
  $a=0.99$ model described in the text. The initial \Bf\ configuration
  corresponds to a neutron star type dipolar field where there are no currents
  outside the event horizon.  The left panel shows the initial
  time, while the right panel shows the solution at
  $t=1000GM/c^3$. The structure of the \ms\ relaxes to monopolar-like
  solution, as predicted by \cite{2011arXiv1104.1091L}. Note also the
  development of the tearing modes and the formation of magnetic
  islands in the equatorial current sheet.}
\end{center}
\label{aphi}
\end{figure}

\begin{figure}[htb]
\begin{center}
\includegraphics[width=0.5\linewidth]{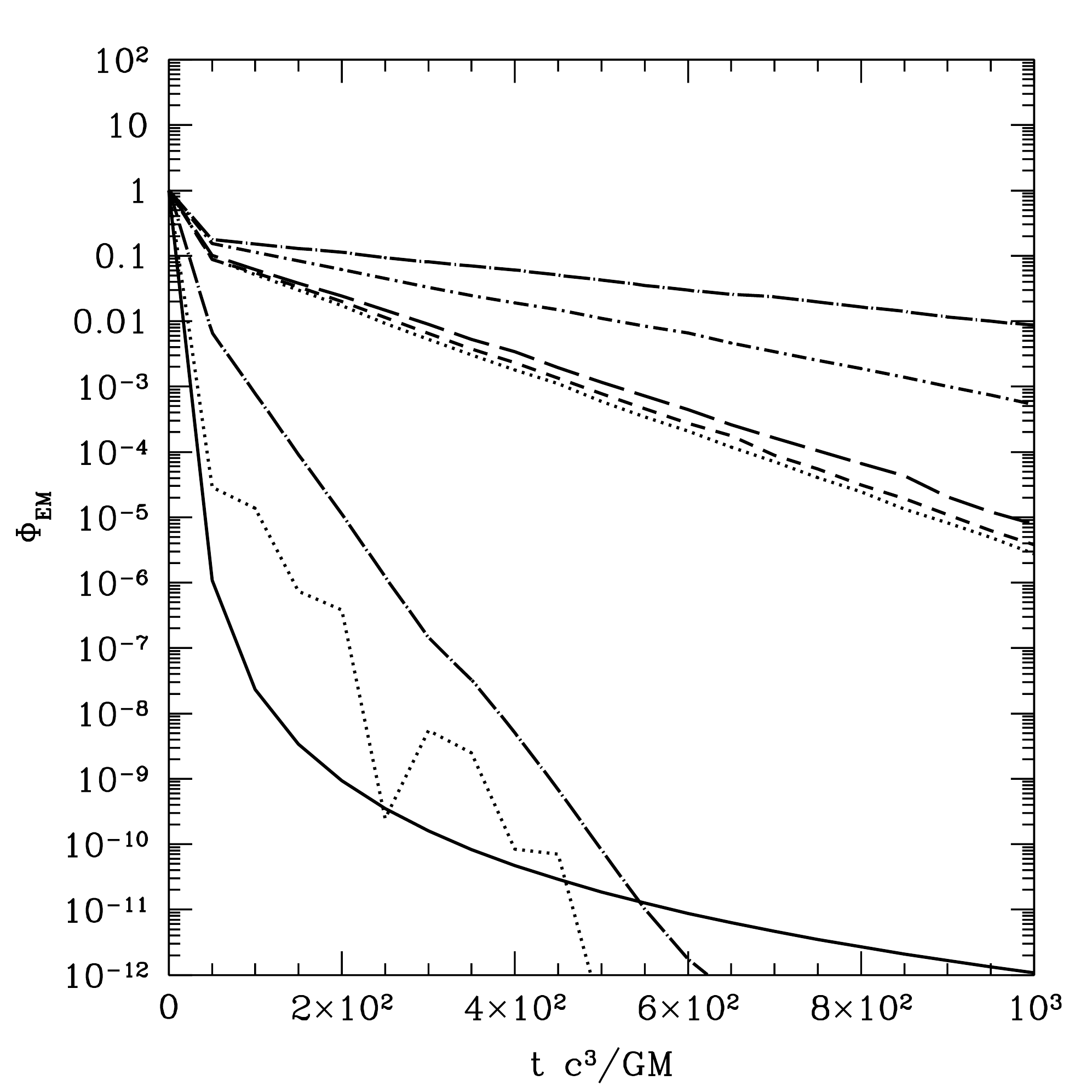}
\caption{ The value of magnetic flux $\Phi_{\rm EM}$ divided by its
  initial ($t=0$) value vs. time ($t$) for the different black hole
  spins and equations of motion (force-free and MHD). The MHD models
  for spins $\{a=0,0.1,0.5,0.9,0.99\}$ correspond to $\{{\rm dot,short
    dash, long dash, dot - short dash, dot - long dash}\}$ line types
  appearing in the upper part of the plot. The force-free models for
  spins $\{a=0,0.99\}$ correspond to $\{{\rm dot,dot - long dash}\}$
  line types and appear at the lower part of the plot. The analytical
  $(t-19)^{-4}$ decay for a vacuum dipole on a black hole is shown as
  a solid line and appears in the lower part of the plot. Notice that
  for the radiatively inefficient MHD regime,
  magnetic flux ``slides off'' the \BH\ on time scales much
  longer than those predicted by the ``no hair'' theorem that follows
  $t^{-4}$.  Also note that the $a=0$ model only mimics a slowly
  rotating NS that still allows a plasma magnetosphere to be self-generated.}
\end{center}
\label{phivst}
\end{figure}

Overall, the field lines that were initially connected to  the surface of
the \NS\ (that would have reached to infinity for a rotating
magnetosphere) remain connected to the black hole horizon for times
much longer than $\sim 20GM/c^3$ that is predicted by the ``no-hair''
theorem.  In case of an initial \Bf\ corresponding to the aligned
pulsar, those field lines that are associated with the closed part of
the magnetosphere are absorbed by the \BH, while those field lines
that would reach to infinity for the \NS\ magnetosphere are forced to
become open by poloidal currents driven into the magnetosphere by the
black hole rotation.

\section{Slowly balding \BH}
 
As we showed above through analytical estimates and numerical
simulations, the \ms\ of a newly formed \BH\ relaxes to a
split-monopole-type structure. The resulting current sheet is subject
to resistive dissipation that would reconnect the field lines from the
different hemispheres, producing a set of closed field lines that will
be quickly absorbed by the \BH\ and a set of open field lines that
will be released to infinity in an event qualitatively similar to
solar coronal mass ejections (CMEs) \citep[][]{2005psci.book.....A}.
This will lead to a decrease of the
number of magnetic flux tubes through each hemisphere $N_B$. The \BH\
will be slowly balding. This ``hair loss'' will proceed on the
resistive time scale of the equatorial current sheet.  Typically,
resistive time scales in a plasma are much longer than the dynamical
times scales by a factor of the so-called Lundquist (magnetic
Reynolds) number.
 
Reconnection of \Bf\ lines is a notoriously difficult problem in
plasma physics \citep[][]{Kulsrud}. Development of plasma turbulence
in the regions of strong current and the resulting anomalous
resistively, plasma collisionality \citep{mu10}, as well as formation
of localized narrow current sheets may bring significant variations
in the dissipation time scale.

As a possible estimate of the reconnection time scale, let us assume
that the reconnection is driven by charge starvation as applicable to
our model.  A charge-neutral plasma of a given density $n$ can support
a current no larger than $j \leq 2 n e c$. Thus, for a given \Bf\ $B$
and density $n$ there is a minimum thickness of a current layer
$\delta \approx B /(2 \pi e n) $.

The \BH\ \ms\ extends from the horizon $R_{\rm BH}$ to the light
cylinder located at $R_{LC}= c/\Omega_H$. General relativistic effects
of time dilation would effectively freeze out the reconnection process
near the horizon. Let us next estimate the reconnection rate near the
light cylinder of the resulting \BH. Taking $\delta$ as a thickness of
the resistive layer of total length $L \sim R _{LC}$, we can estimate
the Lundquist (magnetic Reynolds) number within, \eg the Sweet-Parker
\citep{Kulsrud} model of reconnection, $S=(R_{LC}/\delta)^2$.  In a
highly magnetized plasma the Sweet-Parker reconnection inflow velocity
is $v_{\rm in} \sim c / \sqrt{S}$.

The expected density can be scaled to the local Goldreich-Julian
density \citep{GoldreichJulian}, $n = \lambda n_{GJ}$.  Taking the
angular velocity of field lines equal to $\Omega_H$, then $n_{GJ} = B
\Omega_H/(2 \pi e c)$.  The multiplicity $\lambda$ is expected to be
high, $\lambda \gg 1$ \citep{1975ApJ...196...51R}.  We then find the
Lundquist number $S= 16 \lambda^2 $ and reconnection time scale
$\tau_{\rm rec} = 4 \lambda R_{LC}/c$, which can be written as
\be
{ \tau_{\rm rec} \over R_{\rm BH}/c} \approx  { \lambda \over \chi}   {  R_{\rm BH}  c P_{NS} \over R_{NS}^2} = 10^5 \left({ \chi \over 0.5}\right)^{-1} { \lambda \over 10^5} \, {P_{NS}  \over 1 {\rm msec}} .
\ee
This is much longer than the light travel time over the horizon of the
\BH.

The above estimate of the reconnection rate is given mostly for
illustrative purposes. Various reconnection-type phenomena may produce
vastly different time scales.  For example, the development of a tearing mode
in the relativistic magnetized plasma proceeds on the tearing mode
time scale, which is intermediate between the dynamic and the
resistive time scales \citep{LyutikovTear,2007MNRAS.374..415K}.  In
general, the dissipation timescale is much longer than the dynamical
time scale due to the formation of self-generated currents outside the
event horizon and due to the formation of plasma pressure within the
current sheet.

\section{Discussion}
 
We found that the ``no hair'' theorem is not applicable to a rotating
\NS\ collapsing into a \BH.  As long as the \BH\ is able to
self-produce a highly conducting plasma via the vacuum breakdown, the \Bf\
cannot ``slide off'' the \BH. The presence of a highly conducting
plasma introduces a topological constraint for magnetic fields lines.
If the magnetospheric plasma were ideal, the ``no hair'' theorem would
be truly inapplicable. Only the introduction of a finite resistively,
that leads to a violation of the frozen-in condition, results in loss
of the magnetic field lines.  As a result, a \BH\ retains \Bf\ for
long resistive time scale and not the dynamical time scale $\sim
GM/c^3$ predicted by the ``no hair'' theorem.  Thus, a \BH\ can
produce an \EM\ power for a long time after the collapse, without a
need for an externally supplied \Bf.

The principal difference from the conventional Blandford-Znajek
\cite{Blandford:1977} picture of AGN jet launching is that in that
case the \Bf\ is brought from outside, electric currents are supported
by externally supplied accretion material. In our case there is no
outside-provided plasma: the currents all are self-generated. The
initial poloidal \Bf\ of a NS is necessary to create the \BH\
magnetosphere.  The neutron star's rotation can create poloidal
currents outside what will become the event horizon.  After the
collapse, the initial toroidal currents of the dipole field are
consumed by the BH.  Our simulations show that, even without existing
external poloidal or toroidal currents, the rotation of the newly
formed black hole alone can create new monopolar currents outside the
horizon that can survive for times much longer than the collapse time.

The fact that a \BH\ resulting from a collapse of magnetized rotating
progenitor retains the progenitor's \Bf\ may have important
astrophysical implications, especially in gamma ray burst (GRB)
research, allowing electromagnetic extraction of energy from isolated
\BHs.  The \BH\ resulting from a collapse of a \NS\ will  rotate with angular velocity
\be
\Omega_H \approx {\chi \over 5} { c^4 R^2 \Omega \over G^2  M_{NS}^2} .
\ee
The collapse of the rotating neutron star into the  black hole  preserves the open magnetic flux
\be 
\Phi_0 = \pi R_{NS}^2 B_{NS} \left( {R_{NS} \Omega \over c} \right) .
\ee
The flux $\Phi_0$ will produce magnetic field  on the  black hole 
$
B_{BH} \approx { \Phi_0/(2 \pi R_{BH}^2)}
$.
The spin-down of the resulting magnetized \BH\ will produce an 
\EM\ wind with luminosity  \citep{Michel73,2001MNRAS.326L..41K,mg04,2005ApJ...630L...5M,tch10,tch11}
\be
L_{BH} \approx {2\over 3 c} \left( { \Omega_H\Phi_0 \over 4\pi} \right)^2 =   { 2\pi^4 \over 75} \chi^2 {c^5 \over G^4} { B_{NS}^2  R_{NS}^{10} \over M_{NS}^4 P_{NS}^4} \approx  10^{43}\, {\rm erg s^{-1}} \, \left( { \chi \over 0.5}\right)^2 \,  \left({B_{NS} \over 10^{12} {\rm G}} \right)^2  \, \left( {P_{NS}  \over 1 {\rm msec}}\right)^{-4} .
\ee
This black hole power drives a jet that can reach high Lorentz factors at large radii \citep{mck06jfa,2008MNRAS.385L..28B}.
For the chosen value of the \NS\ \Bf, this is a fairly low power, but
it can become observable if \Bf\ is amplified during the collapse to
magnetar values of $\sim 10^{14}$--$10^{15}$ Gauss \citep{TD93}.  If this indeed happens, the 
rotational power of the \BH\ extracted by \Bfs\ may power the prompt
GRB emission \citep{Usov92} or early afterglows \citep{Lyutikov:2009}.

We would like to thank Scott Hughes, Serguei Komissarov and  Luis Lehner for many insightful comments and
the National Institute for Nuclear Theory for hospitality.

 \bibliographystyle{apsrev}


\end{document}